\documentclass[twoside]{dis09}
\usepackage[latin1]{inputenc}
\usepackage[dvips]{graphicx,epsfig,color}
\usepackage{wrapfig,rotating}
\usepackage{amssymb,amsmath,array}

\begin{document}
\title{EPS09 - Global NLO analysis of nuclear PDFs}

%***********************************************************************
% AUTHORS INFORMATION AREA
%***********************************************************************
\author{Kari J. Eskola$^1$, Hannu Paukkunen$^1$ and Carlos A. Salgado$^3$
%
% Optional short acknowledgment: remove next line if non-needed
%\thanks{This is an optional funding source acknowledgment.}
%
% DO NOT MODIFY THE FOLLOWING '\vspace' ARGUMENT
\vspace{.3cm}\\
%
% Addresses and institutions (remove "1- " in case of a single institution)
1- University of Jyv\"askyl\"a, Department of Physics and \\
Helsinki Institute of Physics, Finland \\
%1- School of First Author - Dept of First Author \\
%Address of First Author's school - Country of First Author's
%school
%
% Remove the next three lines in case of a single institution
\vspace{.1cm}\\
2- Universidade de Santiago de Compostela \\
Departamento de F\'\i sica de Part\'\i culas and IGFAE, Spain\\
%2- School of Second Author - Dept of Second Author \\
%Address of Second Author's school - Country of Second Author's school\\
}
%***********************************************************************
% END OF AUTHORS INFORMATION AREA
%***********************************************************************

\maketitle

\begin{abstract}

In this talk, we present our recent work on next-to-leading order (NLO) nuclear parton distribution functions (nPDFs), which we call EPS09. As an extension to earlier NLO analyses, we complement the deep inelastic scattering and Drell-Yan dilepton data by inclusive midrapidity pion production measurements from RHIC to reduce the otherwize large freedom of the nuclear gluon densities. In addition, our Hessian-type error analysis leading to a collection of nPDF error sets, is the first of its kind among the nPDF analyses.

\end{abstract}

\section{Introduction}

The global analyses of the free nucleon parton distribution functions (PDFs) are based on the asymptotic freedom of QCD, parton evolution and factorization. These features provide the justification for computing hard-processes cross-sections, schematically as
$$
\sigma_{AB\rightarrow h+X} = \sum_{i,j} f_{i}^A(Q^2) \otimes \hat{\sigma}_{ij\rightarrow h+X} \otimes f_{j}^B(Q^2),
$$
where $f_{i}$s denote the scale-dependent PDFs, and $\hat{\sigma}_{ij\rightarrow h+X}$ are perturbatively computable coefficients. The factorization theorem has turned out to work extremely well with more and more different types of data included in the free proton analyses. In the case of bound nucleons factorization is not as well-established, but it has nevertheless proven to provide a very good description \cite{Eskola:1998iy,Eskola:1998df,Eskola:2007my,Eskola:2008ca,Hirai:2007sx,deFlorian:2003qf} of the observed nuclear modifications $\sigma_{\rm bound}/\sigma_{\rm free}$ in deep inelastic scattering (DIS) and Drell-Yan (DY) dilepton production involving nuclear targets. Here, we summarize our recent NLO analysis of the nuclear PDFs and, in particular, their uncertainties \cite{Eskola:2009uj}.

\section{Analysis method and framework}

Our analysis follows very similar pattern as most free proton analyses do:

\vspace{0.2cm}
A. {\bf The PDFs are parametrized at an initial scale $Q_0^2$ imposing the sum rules.} In this work we do not parametrize the absolute PDFs, but rather the $x$ and $A$ dependences of the nuclear modification factors $R_i^A(x,Q_0^2)$ on top of a fixed set of free proton PDFs:
\begin{equation}
f_{i}^A(x,Q^2) \equiv R_{i}^A(x,Q^2) f_{i}^{\rm CTEQ6.1M}(x,Q^2). \nonumber
\end{equation}  
Above $f_{i}^{\rm CTEQ6.1M}(x,Q^2)$ refers to a CTEQ set of the free proton PDFs \cite{Stump:2003yu} in the zero-mass variable flavour number scheme, and we consider three different modification factors: $R_V^A(x,Q_0^2)$ for both $u$ and $d$ valence quarks, $R_S^A(x,Q_0^2)$ for all sea quarks, and $R_G^A(x,Q_0^2)$ for gluons.

\vspace{0.2cm}
B. {\bf The nuclear PDFs are evolved to other perturbative scales $Q^2 > Q_0^2$ by the DGLAP equations.} An efficient numerical solver for the parton evolution is an indispensable ingredient for any parton analysis, but in the case of nuclear PDFs this is even more critical as we need to repeatedly do the evolution separately for 13 different nuclei. On the other hand, the relatively low $Q^2$-values spanned by the data utilized, $Q^2 \lesssim 200 \, {\rm GeV}^2$, and the fact that they come as ratios $\sigma_{\rm bound}/\sigma_{\rm free}$ which are more stable against small evolution inaccuracies, make such task somewhat easier. Our DGLAP code is based on a semi-analytical method described e.g. in \cite{Santorelli:1998yt,Paukkunen:2009ks}.

\vspace{0.2cm}
C. {\bf The cross-sections are computed using the factorization theorem.}

\vspace{0.2cm}
D. {\bf The computed cross-sections are compared with the experimental data, and the initial parametrization is varied to establish an optimal fit to the data.} We define the best agreement as the minimum of a generalized $\chi^2$-function
\begin{eqnarray}
\chi^2(\{a\})    \equiv  \sum _N w_N \, \chi^2_N(\{a\}), \qquad \,\,
\chi^2_N(\{a\})  \equiv  \left( \frac{1-f_N}{\sigma_N^{\rm norm}} \right)^2 + \sum_{i \in N}
\left[\frac{ f_N D_i - T_i(\{a\})}{\sigma_i}\right]^2. \nonumber
\end{eqnarray}
Within each data set $N$, the $D_i$ denotes the experimental data value with $\sigma_i$ point-to-point uncertainty, and $T_i$ is the theory prediction corresponding to a parameter set $\{a\}$. For the pion data, the PHENIX experiment estimates an overall $\sim 10\%$ normalization uncertainty $\sigma_N^{\rm norm}$, and the normalization factor $f_N \in [1-\sigma_N^{\rm norm},1+\sigma_N^{\rm norm}]$ is non-trivial, i.e. $f_N\neq 1$. Its value is determined by minimizing $\chi^2_N$ and the final $f_N$ is thus an output of the analysis. The weight factors $w_N$ are used to amplify the importance of those data sets whose content is physically relevant, but contribution to $\chi^2$ would otherwize be too small to be noticed by an automated minimization.

\vspace{0.2cm}
E. {\bf The uncertainties are estimated.} Besides finding the central set of parameters $\{a^0\}$ that optimally fits the data, quantifying the uncertainties stemming from the experimental errors has become an integral part of the modern PDF fits. In this work, we employ the Hessian method \cite{Pumplin:2001ct}, which rests on a quadratic approximation 
\begin{equation}
\chi^2 \approx \chi^2_0 + \sum_{ij} \frac{1}{2} \frac{\partial^2 \chi^2}{\partial a_i \partial a_j} (a_i-a_i^0)(a_j-a_j^0) \equiv \chi^2_0 +  \sum_{ij} H_{ij}(a_i-a_i^0)(a_j-a_j^0), \nonumber
\end{equation}
around the vicinity of the minimum $\chi^2_0$. 
Non-zero off-diagonal elements in the Hessian matrix $H$ defined above are a signal of correlations between the original fit parameters and it is useful to make a change of variables $\{a\} \rightarrow \{z\}$ that diagonalizes the Hessian matrix. Constructing the so-called PDF error sets $S_k^\pm$ is what ultimately makes the Hessian method so useful. Each $S_k^\pm$ is obtained by displacing the fit parameters to the positive/negative direction along the eigenvector $z_k$ of the Hessian matrix such that $\chi^2$ grows by a certain amount $\Delta \chi^2$. Using these sets, the upper and lower uncertainty of a quantity $X$ can be written e.g. as
\begin{eqnarray}
(\Delta X^+)^2 & \approx & \sum_k \left[ \max\left\{ X(S^+_k)-X(S^0), X(S^-_k)-X(S^0),0 \right\} \right]^2 \label{eq:error_best} \\
(\Delta X^-)^2 & \approx & \sum_k \left[ \max\left\{ X(S^0)-X(S^+_k), X(S^0)-X(S^-_k),0 \right\} \right]^2, \nonumber
\end{eqnarray}
where $X(S^\pm_k)$ denotes the value of the quantity $X$ computed by the set $S_k^\pm$ and where $S^0$ is the best fit. Requiring each data set to remain close to its 90\%-confidence range, we end up with a choice\footnote{See Appendix A of Ref. \cite{Eskola:2009uj} for the detailed explanation.} $\Delta \chi^2=50$. 

\section{Results and Conclusions}

\begin{wrapfigure}{r}{0.7\columnwidth}
\centerline{\includegraphics[width=0.7\columnwidth]{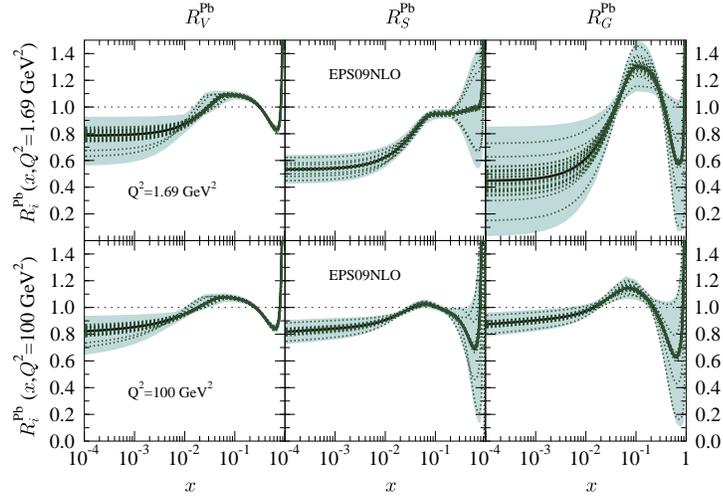}}
\caption{The obtained nuclear modifications for Lead at the initial scale $Q^2_0=1.69 \, {\rm GeV}^2$ and at $Q^2=100 \, {\rm GeV}^2$. The thick black lines indicate the best-fit, whereas the dotted green curves denote the individual error sets which combine to the shaded bands like in Eq.~(\ref{eq:error_best}).}
\label{Fig:PbPDFs}
\end{wrapfigure}
We briefly go through the main results from the present analysis, starting with Fig.~\ref{Fig:PbPDFs} where we plot the obtained modifications for Lead at two scales making their scale-dependence thereby clearly visible. It should be noticed that even a rather large uncertainty band at small-$x$ gluons shrinks in the scale evolution --- a clear prediction of the DGLAP approach that might be testable in the future colliders. 

\begin{figure}[!htb]
\center
\includegraphics[scale=0.27]{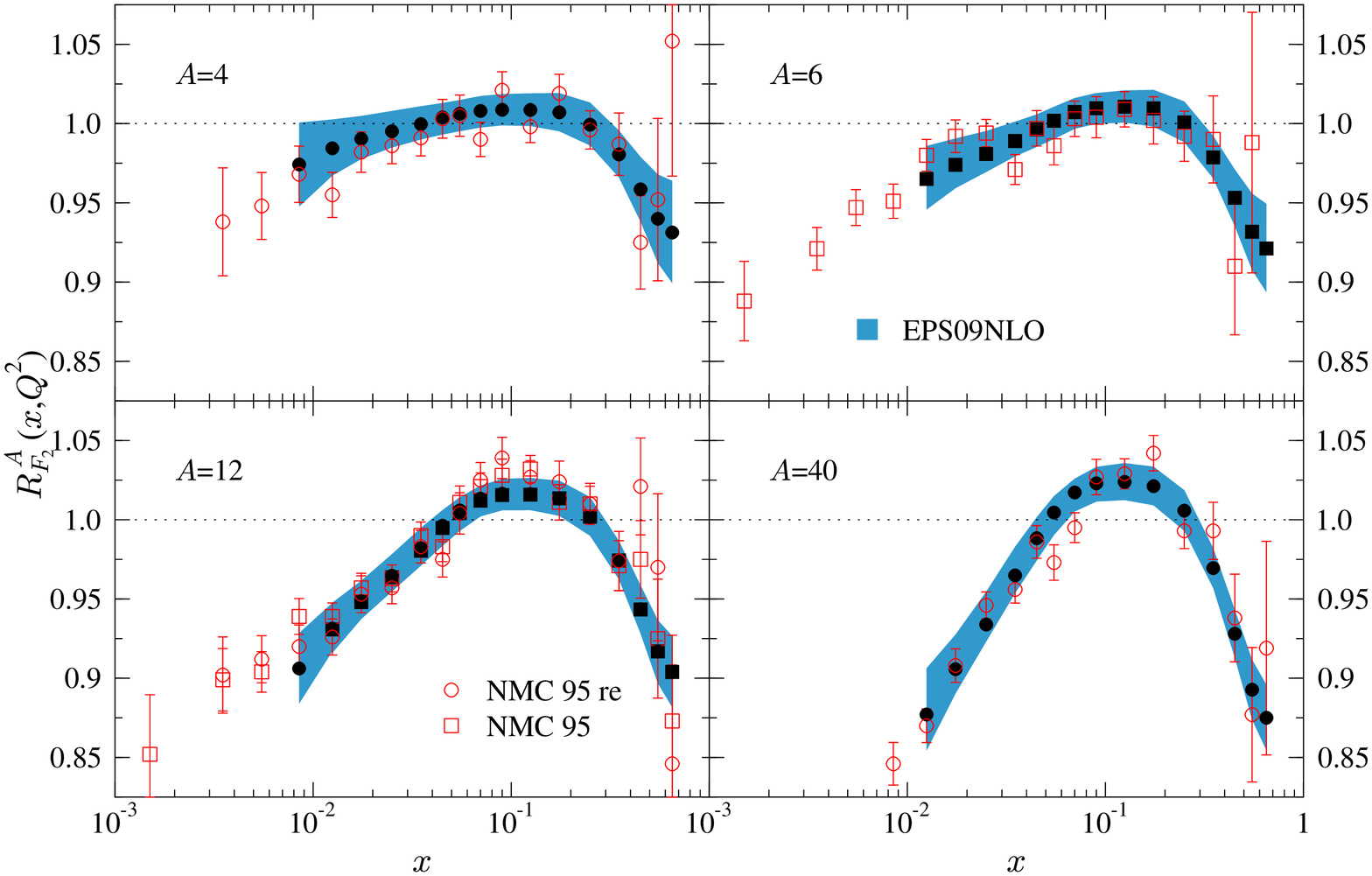}
\includegraphics[scale=0.285]{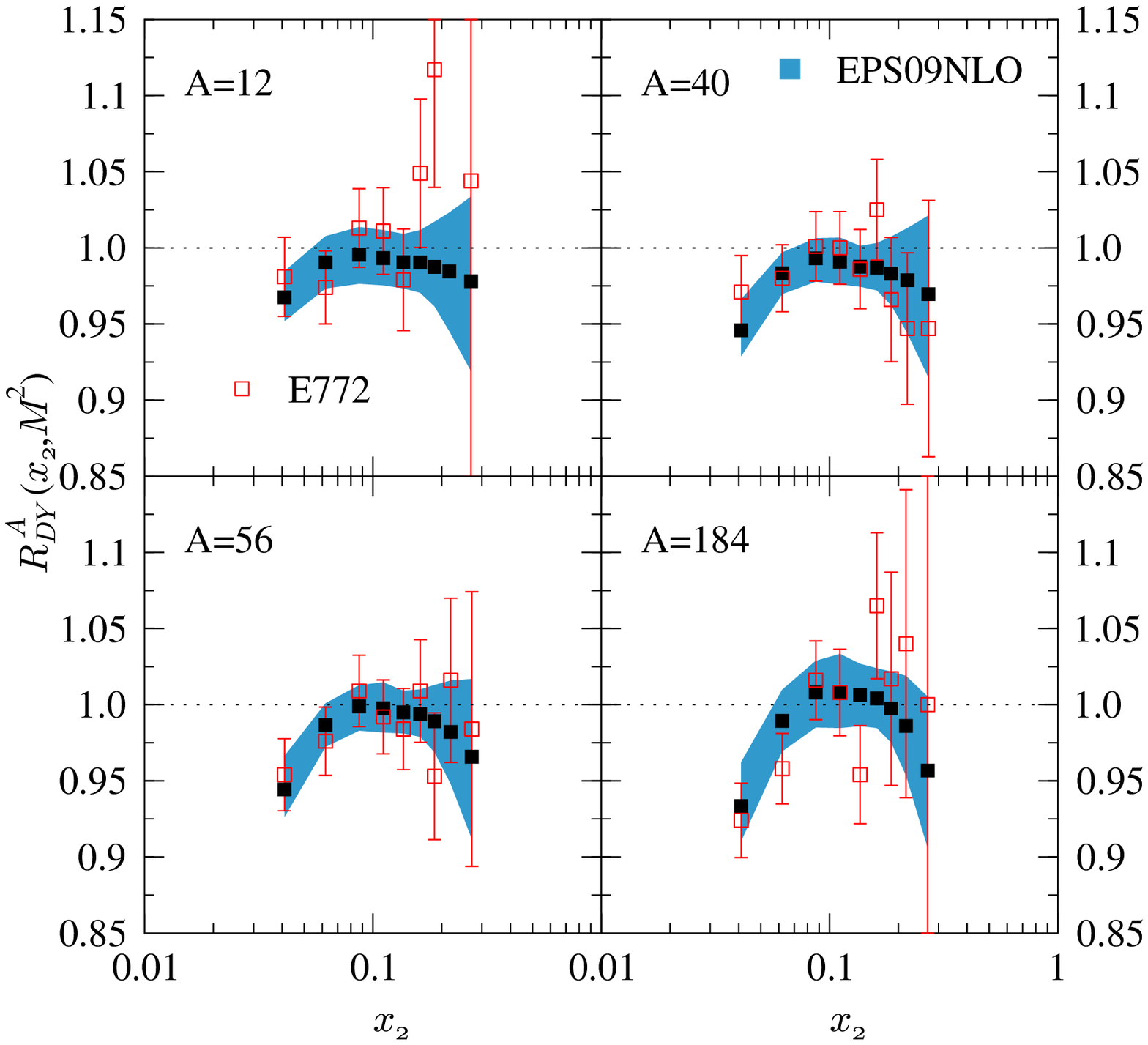}
\caption[]{The calculated $R_{F_2}^A$ and $R_{\rm DY}^{\rm A}$ compared with the NMC \cite{Arneodo:1995cs,Amaudruz:1995tq} and E772 \cite{Alde:1990im} data.}
\label{Fig:RF2A1}
\end{figure}

As the DIS and DY data constitute the major part of the available experimental data we display in Fig.~\ref{Fig:RF2A1} some representative examples of the measured nuclear modifications with respect to Deuterium,
$$
R_{F_2}^{\rm A}(x,Q^2) \equiv  \frac{F_2^A(x,Q^2)}{F_2^d(x,Q^2)},\qquad 
 R_{\rm DY}^{\rm A}(x_{2},M^2) \equiv \frac{\frac{1}{A}d\sigma^{\rm pA}_{\rm DY}/dM^2dx_{2}}{\frac{1}{2}d\sigma^{\rm pd}_{\rm DY}/dM^2dx_{2}}_{\Big|x_{2} \equiv \sqrt{M^2/s}\,e^{-y}}
$$
for different nuclei compared with the EPS09.
The shaded blue bands denote the uncertainty propagated from the 30 EPS09 error sets and, as should be emphasized, their size is comparable to the experimental errors backing up our choice for the $\Delta \chi^2$. 
\begin{figure}[!htb]
\centering
\includegraphics[scale=0.25]{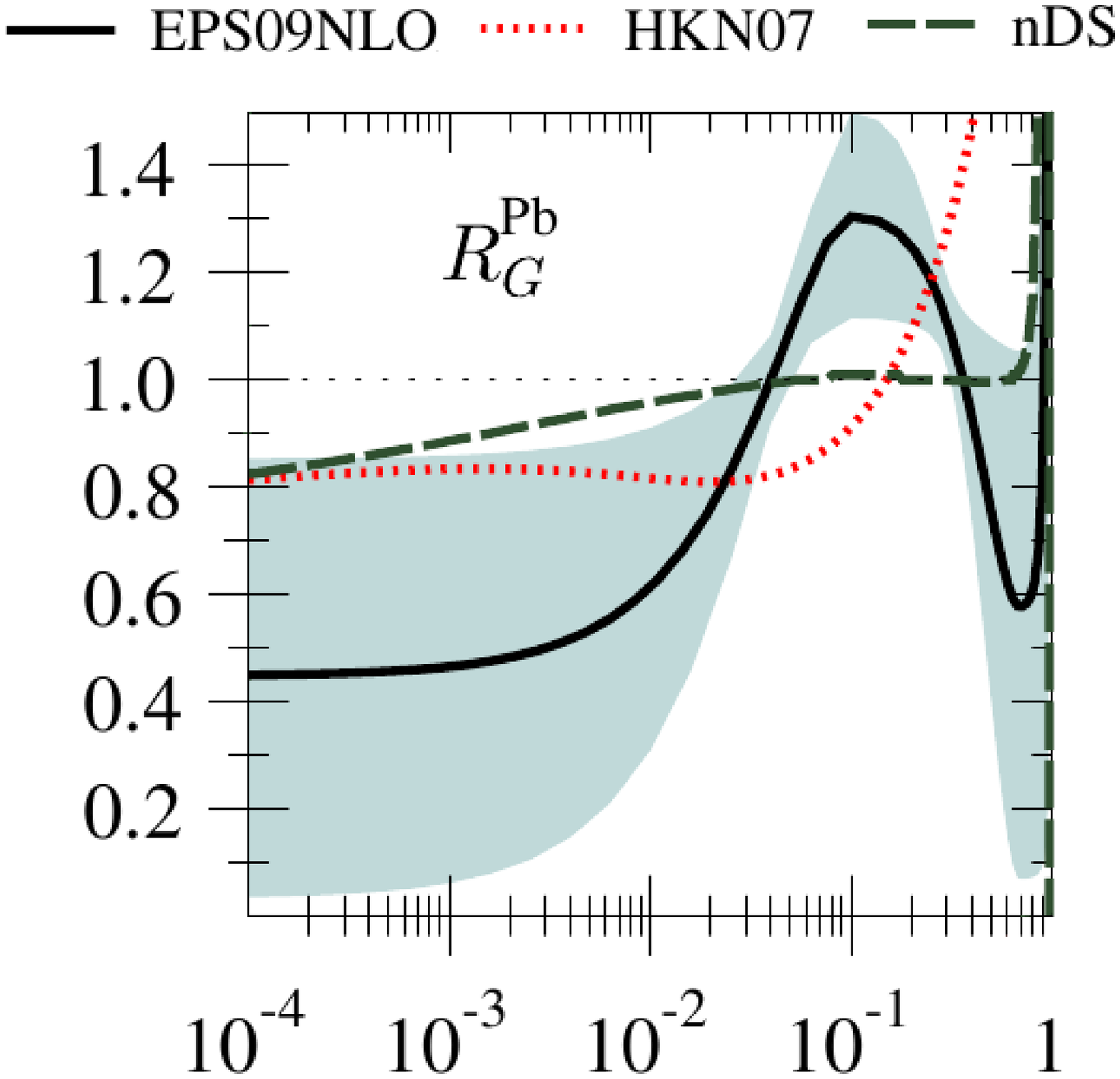}
\includegraphics[scale=0.43]{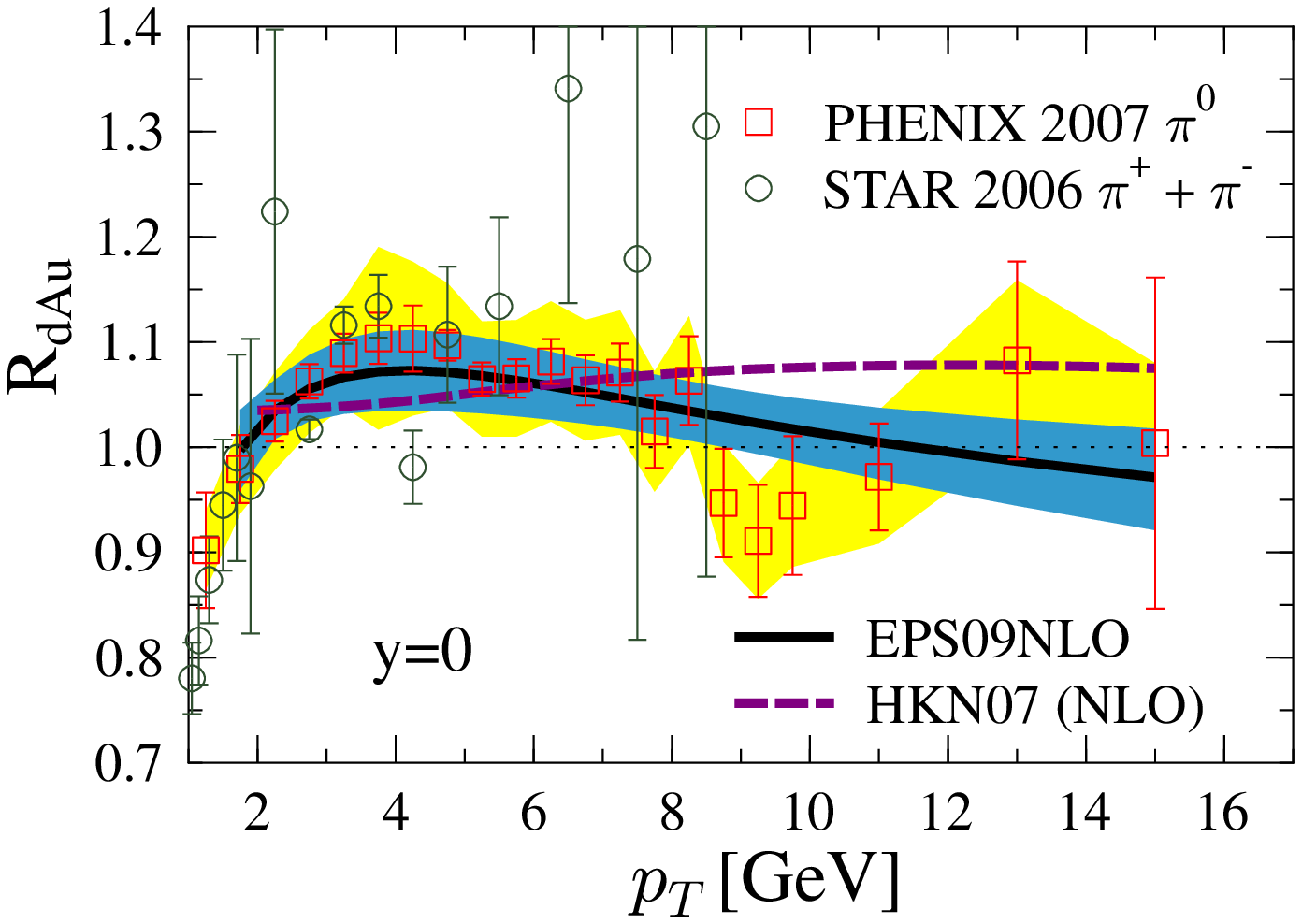}
\caption{Left: Comparison of the nuclear modifications $R_G^{\rm Pb}$ at $Q^2 = 1.69 \, {\rm GeV}^2$ from HKN07~\cite{Hirai:2007sx}, nDS~\cite{deFlorian:2003qf} and this work, EPS09 \cite{Eskola:2009uj}. Right: The computed $R_{\rm dAu}$ for inclusive pion production compared with the PHENIX \cite{Adler:2006wg} and STAR \cite{Adams:2006nd} data multiplied by $f_N = 1.03$ $f_N = 0.90$, respectively.}
\label{Fig:PHENIX}
\end{figure}
The nuclear modification for inclusive pion yield is defined as
\begin{equation}
R_{\rm dAu}^{\pi}  \equiv  \frac{1}{\langle N_{\rm coll}\rangle} \frac{d^2 N_{\pi}^{\rm dAu}/dp_T dy}{d^2 N_{\pi}^{\rm pp}/dp_T dy} \stackrel{\rm min. bias}{=} \frac{\frac{1}{2A} d^2\sigma_{\pi}^{\rm dAu}/dp_T dy}{d^2\sigma_{\pi}^{\rm pp}/dp_T dy}, \nonumber 
\end{equation}
where $\langle N_{\rm coll}\rangle$ denotes the number of binary nucleon-nucleon collisions and $p_T,y$ the pion's transverse momentum and rapidity. A comparison with the PHENIX and STAR data is shown in Fig.~\ref{Fig:PHENIX}. Evidently, the shape of the spectrum --- which in our calculation is a reflection of the similar shape in $R_G$ --- gets well reproduced by EPS09. Let us mention that the shape is practically independent of the fragmentation functions used in the calculation --- modern sets like \cite{Kniehl:2000fe,Albino:2008fy,deFlorian:2007aj} all give equal results.

Figure~\ref{Fig:PHENIX} also presents a comparison of EPS09 gluon modifications $R_G^{\rm Pb}$ with the earlier NLO analyses. The significant scatter of the curves highlight the difficulty of pinning down the nuclear modifications from the DIS and DY data alone --- especially the behaviour of HKN07 looks different. Consequently, also the predictions for pion $R_{\rm dAu}$ differ significantly as is easily seen in Fig.~\ref{Fig:PHENIX}. This is actually good news as this type of data, especially with better statistics, may eventually discriminate between different proposed gluon modifications. 

\begin{figure}[!htb]
\center
\includegraphics[scale=0.25]{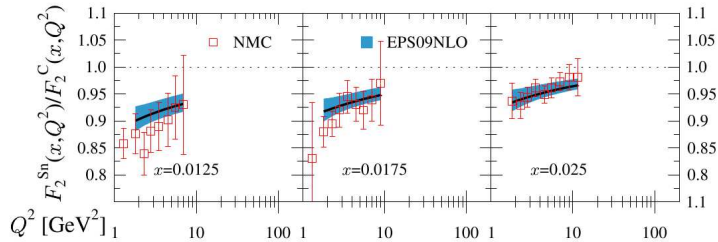}
\caption[]{\small The calculated scale evolution of the ratio $F_2^{\mathrm{Sn}}/F_2^{\mathrm{C}}$ compared with the NMC data \cite{Arneodo:1996ru}.}
\label{Fig:RF2_slopes}
\end{figure}

Attention should be paid to the experimentally observed scaling-violations and to the fact that the DGLAP dynamics reproduces them well. Most cleanly such effects are visible e.g from the small-$x$ structure function ratios versus $Q^2$, of which Fig.~\ref{Fig:RF2_slopes} shows an example.

In summary, the very good agreement with the experimental data $\chi^2/N \approx 0.79$ found --- especially the correct description of the scale-breaking effects --- we argue, is compelling evidence for the applicability of collinear factorization in nuclear environment. In addition to the best fit, we release \cite{EPS09code} 30 nPDF error-sets for practical use, encoding the neighborhood of the $\chi^2$ minimum. Although not discussed here, we have also performed the leading-order counterpart of the NLO analysis as we want to provide the uncertainty tools also for this widely-used framework. Although the best-fit quality is very similar both in LO and NLO, the uncertainty bands become smaller when going to higher order.

% ****************************************************************************
% BIBLIOGRAPHY AREA
% ****************************************************************************

\begin{footnotesize}
% IF YOU DO NOT USE BIBTEX, USE THE FOLLOWING SAMPLE SCHEME FOR THE REFERENCES
% ----------------------------------------------------------------------------

\end{footnotesize}

% ****************************************************************************
% END OF BIBLIOGRAPHY AREA
% ****************************************************************************

\end{document}